# Fraunhofer diffraction of a Laguerre-Gaussian laser beam by fork-shaped grating

Suzana Topuzoski[*] and Ljiljana Janicijevic

*Institute of physics, Faculty of natural sciences and mathematics, Skopje, R. Macedonia*

*Corresponding author: suzana_topuzoski@yahoo.com; suzanat@pmf.ukim.mk*

**Abstract:** In this article we present a theoretical study for Fraunhofer diffraction of a Laguerre-Gaussian laser beam with zeroth radial mode number and azimuthal mode number $l$ by diffractive grating with embedded fork-shaped dislocations of integer order $p$. Analytical expressions describing the diffracted wavefield amplitude and intensity distributions in the Fourier plane are deduced and analyzed. They are also followed by the vortex radii expressions.

## 1. Introduction

Diffractive optical elements (DOEs) which generate laser beams with phase singularities in their wavefronts have been research objects throughout recent years. The wavefront dislocations or topological defects represent discontinuities of the wave phase. There are two pure dislocations [1]: edge dislocation, located along a line in the transverse plane and traveling with the beam, and vortex or screw dislocation, which is characterized with spiral (helicoidal) wave phasefront, rotating as a screw around the line of the dislocation. Along the vortex dislocation line the wave intensity has zeroth value and nondefined phase. Among the DOEs that generate optical vortices are phase spiral plates [2, 3], helical axicons [4, 5, 6] and spiral zone plates [7]. The ability of the fork-shaped holograms and diffraction gratings to produce vortex beams from an incident Gaussian laser beam was experimentally verified [8, 9] and after, widely used for guiding cold atoms [10, 11], optical manipulation of micron-sized objects [12, 13], and in quantum information applications [14, 15], since these vortex beams carry definite photon orbital angular momentum [16].

In [17] the complete analytical solution for the problem of Fraunhofer diffraction and Fresnel diffraction of a Gaussian beam incident out of waist by fork-shaped grating with integer topological charge $p$, is reported. Analytical expressions for the wave amplitude and intensity distributions are derived, describing their radial part in the $m$-th diffraction order by the product of the $mp$-th-order Gauss-doughnut function and a confluent hypergeometric (or Kummer) function, or by the product of the first-order Gauss-doughnut function and the difference of two modified Bessel functions whose orders do not match the singularity number.

A new family of paraxial laser beams that form an orthogonal basis is named as hypergeometric (HyG) modes by the authors in [18], since they have the complex amplitude proportional to the confluent hypergeometric function. Unlike those, the hypergeometric-Gaussian modes carry a finite power and have been generated in [19] with a liquid-crystal spatial light modulator.

Transformation of an incident beam carrying topological charge, like Laguerre-Gaussian (LG) beam with nonzeroth azimuthal mode number, or higher-order Bessel beam, by the DOE with embedded phase singularity is also very interesting. In [20] the problem of Fresnel diffraction of an incident, nondiverging, vortex Bessel beam, having topological charge $n$, by forked grating with integer topological charge $p$, has been solved analytically: the diffracted wave field amplitude in the higher ($m$-th) diffraction order is described as a sum of Gauss hypergeometric functions, and was shown to carry topological charge $(n \pm mp)$, respectively for the positive and negative $m$-th diffraction order.

While, in [21] and [22], respectively, transformation of $\mathrm{LG}_{n=0}^{(l)}$ and $\mathrm{LG}_{n\neq 0}^{(l)}$ beam, into diverging or nondiverging Bessel beam, which can have increased, decreased, or zeroth topological charge number, by means of a helical axicon has been shown.

In [23] the Fresnel diffraction of a Laguerre-Gaussian beam with zeroth radial mode number and arbitrary azimuthal index by forked grating has been treated. The incident beam propagation axis is orthogonal to the diffraction grating plane and intersects it exactly in the centre of the grating bifurcation point, while the beam waist is in the grating plane.

When the problem of diffraction of a specific laser beam by a diffraction grating is considered, the solution is usually considered as a complete when both, the Fresnel diffraction and the Fraunhofer diffraction are treated. In that means, here we refer to the Fresnel diffraction of $\mathrm{LG}_0^{(l)}$ beam by the forked grating when incident with its waist on the grating plane [23] and when incident out of waist on the grating plane (the main results arising from [24] are given here). While, in this paper we treat theoretically for the first time (as far as we know) the problem of Fraunhofer diffraction of $\mathrm{LG}_0^{(l)}$ beam by the forked grating with arbitrary integer topological charge $p$ (including the special case when $p=0$, i.e. the grating is an ordinary rectilinear one, and the case when $l=0$, $p \neq 0$). Moreover, analytical formulas for the vortex radii are derived, what hasn't been done before for this problem, and which is of importance in the experiments for optical manipulation of micron sized objects and atom trapping and guiding. Namely, the holographic optical tweezers use a computer designed DOE to split a single collimated laser beam into several separate beams, each of which is focused into an optical tweezers by a strongly converging lens [25].

## 2. Fraunhofer diffraction of $\mathrm{LG}_0^{(l)}$ beam by the fork-shaped grating

The incident Laguerre-Gaussian beam is with radial mode number $n=0$ and azimuthal mode number $l$ (taken with positive value), and has its waist in the forked grating plane $\Delta(r,\varphi)$

$$U_0^{(l)}(r,\varphi,\zeta=0) = A_{l,0}\left(\frac{r\sqrt{2}}{w_0}\right)^l \exp\left(-\frac{r^2}{w_0^2}\right)\exp(il\varphi), \qquad (1)$$

where the amplitude coefficient is $A_{l,0} = 2\sqrt{1/(1+\delta_{0,l})\pi l!}$, $k = 2\pi/\lambda$ is propagation constant, $w_0$ is the beam waist radius.

We treat the problem of Fraunhofer diffraction of the beam (1) by fork-shaped grating (FSG), having transmission function

$$T(r,\varphi) = \sum_{m=-\infty}^{\infty} t_m \exp\left[-im\left(\frac{2\pi}{D}r\cos\varphi - p\varphi\right)\right], \qquad (2)$$

where $p$ is an integer topological charge order of the grating, showing the number of its internal "teeth". When $p=0$ this transmission function describes a rectilinear grating. $D$ is period of the rectilinear grating and plays the same role for the FSG far from its pole. While, the specifications of the transmission coefficients $t_m$ depend on the type of the grating, and we define them as in [17], for the cases of amplitude hologram and amplitude binary grating, and their phase versions. The incident beam is entering normally to the grating plane $\Delta(r,\varphi)$, passing with its axis through the pole of the grating. We calculate the diffracted wave field amplitude in the focal plane $\Pi(\rho,\theta)$ of a convergent lens with focal distance $f$, using the diffraction integral

$$U(\rho,\theta,f) = C\iint_\sigma T(r,\varphi)U_0^{(l)}(r,\varphi)\exp\left(\frac{ik\rho}{f}r\cos(\varphi-\theta)\right)r\,dr\,d\varphi\;,$$

in which we involve the grating transmission function (2) and the incident beam expression (1). In the upper integral $C = i/\lambda f$ is a complex constant, while with $\sigma$ the grating area which contributes to the diffraction is denoted. The integration over the angular variable is performed by means of introducing new variables $(\rho_m,\theta_m)$ and $(\rho_{-m},\theta_{-m})$ in the observation plane [17]

$$\rho_{\pm m} = \sqrt{\rho^2 + \left[\frac{m\lambda f}{D}\right]^2 \mp \frac{2m\lambda\rho f}{D}\cos\theta}; \quad \tan\theta_{\pm m} = \frac{\rho\sin\theta}{\rho\cos\theta \mp m\lambda f/D}.$$

The performed integration over the azimuthal variable $\varphi$ leads to the following expression

$$U(\rho,\theta,f) = 2\pi C A_{l,0}\left(\frac{\sqrt{2}}{w_0}\right)^l \Big\{ t_0 \exp[il(\theta+\pi/2)]Y_0(\rho) + \sum_{m=1}^{\infty} t_{+m}\exp[is_{+m}(\theta_{+m}+\pi/2)]Y_{+m}(\rho_{+m})$$
$$+ \sum_{m=1}^{\infty} t_{-m}\exp[is_{-m}(\theta_{-m}+\pi/2)]Y_{-m}(\rho_{-m}) \Big\}. \qquad (3)$$

In (3) the integrals over the radial variable are denoted as

$$Y_0(\rho) = \int_0^\infty \exp\left(-\frac{r^2}{w_0^2}\right)J_l\left(\frac{k\rho r}{f}\right)r^{l+1}dr\;;\quad Y_{\pm m}(\rho_{\pm m}) = \int_0^\infty \exp\left(-\frac{r^2}{w_0^2}\right)J_{|s_{\pm m}|}\left(\frac{k\rho_{\pm m}r}{f}\right)r^{l+1}dr,$$

and the following signs are introduced

$$s_{+m} = l + mp; \quad s_{-m} = l - mp; \quad (m=1,2,...) ; \text{ and } s_{m=0} = l. \tag{4}$$

The upper integrals are the well-known integrals of Bessel functions [26], used in [17]. Their solutions lead to the final analytical form for the diffracted wave field, which we write as a sum of a zeroth-diffraction-order beam and higher, positive and negative diffraction-order beams

$$U(\rho,\theta,f) = U_0(\rho,\theta,f) + \sum_{m=1}^{\infty} U_{+m}(\rho_{+m},\theta_{+m},f) + \sum_{m=1}^{\infty} U_{-m}(\rho_{-m},\theta_{-m},f). \tag{5}$$

The diffracted beam in the zeroth diffraction order is found as LG beam with a phase singularity $l$ along its propagation axis ($\rho=0$), where it has zeroth value amplitude

$$U_0(\rho,\theta,f) = A_{l,0} i \frac{w_0}{w_f} t_0(\sqrt{2})^l \exp[il(\theta + \pi/2)] \left(\frac{\rho}{w_f}\right)^l \exp\left(-\frac{\rho^2}{w_f^2}\right). \tag{6}$$

In (6) the notation for $w_f = \lambda f / w_0 \pi$ has been used.

The beams in the higher diffraction orders, positive and negative, are deviated from the $z$-axis, on the right and left side, for angles $\delta_{\pm m} = \arctan(m\lambda/D)$. Their wave amplitudes are found in the form

$$U_{\pm m}(\rho_{\pm m},\theta_{\pm m},f) = A_{l,0} i \frac{w_0}{w_f} t_{\pm m}(\sqrt{2})^l \exp[is_{\pm m}(\theta_{\pm m} \pm \pi/2)]$$

$$\times \left(\frac{\rho_{\pm m}}{w_f}\right)^{|s_{\pm m}|} \exp\left(-\frac{\rho_{\pm m}^2}{w_f^2}\right) \frac{\Gamma((|s_{\pm m}|+l)/2+1)}{\Gamma(|s_{\pm m}|+1)} M\left(\frac{|s_{\pm m}|-l}{2}, |s_{\pm m}|+1; \frac{\rho_{\pm m}^2}{w_f^2}\right). \tag{7}$$

From the expressions (7) it can be concluded that, in the ($\pm m$)-th diffraction order, a phase singularity of $|s_{\pm m}|$-th order occurs. The topological charge is equal to $l+mp$ and $l-mp$, respectively, for the positive and negative diffraction-order beams. The radial part of the wave functions is represented as a product of a Gauss-doughnut function of order $|s_{\pm m}|$, $(\rho_{\pm m}/w_f)^{|s_{\pm m}|} \exp(-\rho_{\pm m}^2/w_f^2)$, and a confluent hypergeometric (Kummer) function of real argument, $M((|s_{\pm m}|-l)/2, |s_{\pm m}|+1; \rho_{\pm m}^2/w_f^2)$.

Further, we'll analyze the change of the beams topological charge for different correlation between $l$ and $p$:

**a)** $l<p$ ($s_{-m} < 0$) ($l > 0, p > 0$)

The zeroth-diffraction-order beam has phase singularity order and helicity of its wavefront as those of the incident beam. In the positive diffraction orders, the topological charge is increasing as $l+mp$, and the wavefronts helicity is in the same rotation direction as for the incident beam. Whereas, in the negative diffraction orders, the wavefronts helicity has an opposite rotation direction, compared to that of the incident beam (the topological charge $l-mp$ has negative sign), while, the singularity order increases when going to the right, towards the higher negative diffraction orders (an example is shown in Figure 1).

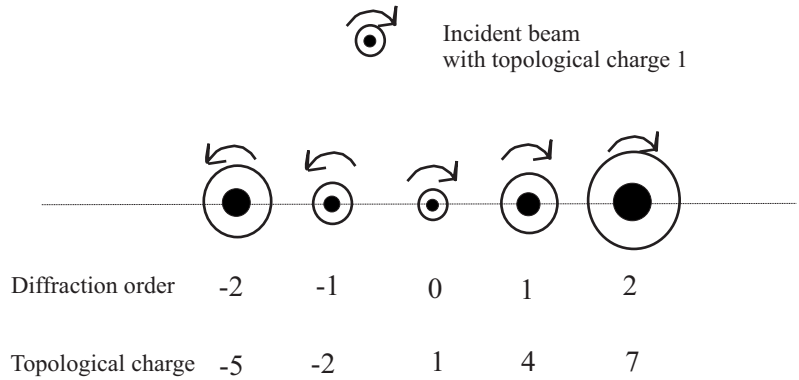

Figure 1. Change of the incident LG beam topological charge in the diffraction orders, for the case: *l*=1, *p*=3 (radial number *n*=1). The arrows show the direction of the phase wavefront helicity of the beams in the coresponding diffraction orders.

**b)** $|l| \geq |p|$ ($l > 0, p > 0$)

The zeroth-diffraction-order beam has phase singularity order *l* and helicity of its wavefront same as of the incident beam.

The positive-diffraction-order beams have topological charges *l+mp*, and their wavefronts helicity is in the same direction as for the incident beam. In a given negative diffraction order $m_0$ it might be satisfied: $l - m_0 p = 0$, or: $m_0 = l/p$. In this diffraction order, then, the beam is without phase singularity, having a non-zero wave amplitude along its propagation axis. This place, now, will act as a "referent place"–on its right side the wavefronts helicity has an opposite direction from that of the incident beam, and the topological charge is equal to *l-mp*. Whereas, on the left side from this "referent place", the wavefronts helicity is in the same direction as for the incident beam (see example in Figure 2).

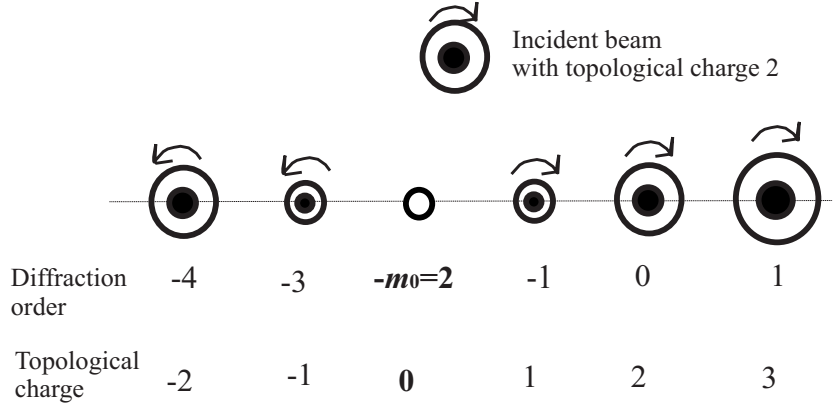

Figure 2. Change of the incident LG beam topological charge for the case $l=2$, $p=1$ ($n=1$).

It is interesting to notice the case when: $l - m_0 p = 0$. Then, from the Equation (7) we get the amplitude distribution in the negative $m_0$-th diffraction order as follows

$$U_{-m_0}(\rho_{-m_0}, \theta_{-m_0}, f) = iA_{l,0} t_{-m_0} (\sqrt{2})^l \frac{w_0}{w_f} \Gamma(l/2+1) \exp\left(-\frac{\rho_{-m_0}^2}{w_f^2}\right) M\left(-l/2, 1; \frac{\rho_{-m_0}^2}{w_f^2}\right). \quad (8)$$

Its intensity distribution along the propagation axis is different from zero

$$I_{-m_0}(\rho_{-m_0}=0, \theta_{-m_0}, z) \propto (w_0/w_f)^2 2^l \Gamma^2(l/2+1) M^2(-l/2, 1; 0) \neq 0. \quad (9)$$

If it's satisfied: $l - m_0 p = 0$ and $l$ is an even number at the same time: $l=2n'$ ($n'$ is an integer), then, from the solution (8), using the relation between Laguerre polynomial and Kummer function [26]

$$M(-n', \alpha+1; x) = \frac{(n')!}{(\alpha+1)_{n'}} L_{n'}^{(\alpha)}(x),$$

we get the diffracted wave amplitude in this negative diffraction order $m_0$ in the form

$$U_{-m_0}(\rho_{-m_0}, \theta_{-m_0}, z) = iA_{2n',0} t_{-m_0} 2^{n'} [(n')!]^2 \frac{w_0}{w_f} \exp\left(-\frac{\rho_{-m_0}^2}{w_f^2}\right) L_{n'}^{(0)}\left(\frac{\rho_{-m_0}^2}{w_f^2}\right). \quad (10)$$

One can see that the beam given by Equation (10) is a pure LG mode, with radial mode number $n'$ and azimuthal mode number equal to zero.

The beams diffracted in the higher diffraction orders, positive and negative, have phase singularities $s_{\pm m} = l \pm mp$ of the variables $\theta_{\pm m}$, in the points $C_{+m}(m\lambda f/D, 0)$ and $C_{-m}(m\lambda f/D, \pi)$, where $\rho_{\pm m} = 0$. The topological charge in these diffraction orders is changing in the way explained in Figures 1 and 2.

The expressions (7) are similar to the far-field approximations of the higher-diffraction-order amplitudes, obtained in the process of Fresnel diffraction of an out of waist incident $LG_{n=0}^{(l)}$ beam by forked grating, derived and discussed in detailed in [24], and to the results presented in [23]. In [24] the incident $\mathrm{LG}_0^{(l)}$ beam enters into the forked grating plane normally, but, differing from [23], its waist is shifted a distance $z=\zeta$ from the grating. The higher-diffraction-order beams in the far-field approximation are found as [24]

$$U_{\pm m}(\rho_{\pm m},\theta_{\pm m},z) \approx A_{l,0} t_{\pm m} (\sqrt{2})^l \frac{w_0}{w(z)} (w(\zeta)w(z))^{\frac{|s_{\pm m}|-l}{2}} \left(\frac{i\pi}{\lambda(z-\zeta)}\right)^{\frac{|s_{\pm m}|-l}{2}} \frac{\Gamma((l+|s_{\pm m}|)/2+1)}{\Gamma(|s_{\pm m}|+1)}$$

$$\times \exp(-i\phi''_g) \exp[is_{\pm m}(\theta_{\pm m}+\pi/2)] \left(\frac{\rho_{\pm m}}{w(z)}\right)^{|s_{\pm m}|} \exp\left(-\frac{\rho_{\pm m}^2}{w^2(z)}\right) M\left((|s_{\pm m}|-l)/2, |s_{\pm m}|+1; \frac{\rho_{\pm m}^2}{w^2(z)}\right), \quad (11)$$

where $\phi''_g \approx kz - \left(\frac{l-|s_{\pm m}|}{2}\right) \arctan\left(\frac{\zeta}{z_0}\right) - \left(\frac{l+|s_{\pm m}|}{2}+1\right) \arctan\left(\frac{z}{z_0}\right)$ is the Guoy phase.

In the expressions (7) the variables $w_0$ and $w_f$ are present, instead of $w(\zeta)$ and $w(z)$, respectively. $w(z) = w_0 [1+(z/z_0)^2]^{1/2}$ is the beam transverse amplitude profile radius for the fundamental (Gaussian) mode, at distance $z$ from its beam waist $w_0$, $z_0 = kw_0^2/2$ is the Rayleigh distance. These far-field approximate expressions arise from the general solution of the Fresnel diffraction of $\mathrm{LG}_0^{(l)}$ beam incident with its waist a distance $z=\zeta$ from the forked grating plane [24]

$$U_{\pm m}(\rho_{\pm m},\theta_{\pm m},z) = A_{l,0} t_{\pm m} i^{|s_{\pm m}|} \left(\frac{\sqrt{2}}{w(\zeta)}\right)^l \frac{w_0}{w(z)} \left[\frac{w(\zeta)}{w(z)}\right]^{\frac{|s_{\pm m}|+l}{2}} \left[\frac{ik}{2(z-\zeta)}\right]^{\frac{|s_{\pm m}|-l}{2}} \frac{\Gamma((l+|s_{\pm m}|)/2+1)}{\Gamma(|s_{\pm m}|+1)}$$

$$\times \exp(-i\phi_g) \exp[is_{\pm m}(\theta_{\pm m}+\pi/2)] \exp\left(-\frac{\rho_{\pm m}^2}{w^2(z)}\right) \rho_{\pm m}^{|s_{\pm m}|} M\left((|s_{\pm m}|-l)/2, |s_{\pm m}|+1; -\frac{ikq(\zeta)}{2q(z)(z-\zeta)} \rho_{\pm m}^2\right).$$
(12)

Now, the Guoy phase is:

$$\phi_g = k\left[z + \frac{\rho_{\pm m}^2}{2}\left(\frac{1}{R(z)} - \frac{1}{z-\zeta}\right) + \frac{\rho^2}{2(z-\zeta)}\right] - \left(\frac{l-|s_{\pm m}|}{2}\right) \arctan\left(\frac{\zeta}{z_0}\right) - \left(\frac{l+|s_{\pm m}|}{2}+1\right) \arctan\left(\frac{z}{z_0}\right),$$

and, with $q(z) = z + ikw_0^2/2$ the beam complex parameter is signed.

## 3. The diffracted intensity distributions and vortex radii in the Fourier plane

In the transverse profile, the central dark spot of the zeroth-diffraction-order beam is surrounded by a bright ring, whose radius is derived using the intensity distribution of the beam (6)

$$I_0(\rho,\theta,f) = |A_{l,0} t_0|^2 (w_0/w_f)^2 2^l (\rho/w_f)^{2l} \exp(-2\rho^2/w_f^2),$$

by searching for its first extreme upon the radial variable $\rho$, as

$$\rho_{0,\max} = w_f \sqrt{l/2}. \qquad (13)$$

The analytical expressions for the vortex radii in the higher-diffraction-order beams can be derived from their intensity distributions

$$I_{\pm m}(\rho_{\pm m}, \theta_{\pm m}, f) = |A_{l,0} t_{\pm m}|^2 \left(\frac{w_0}{w_f}\right)^2 2^l \left(\frac{\rho_{\pm m}^2}{w_f^2}\right)^{|s_{\pm m}|} \exp\left(-\frac{2\rho_{\pm m}^2}{w_f^2}\right)$$
$$\times \frac{\Gamma^2((|s_{\pm m}|+l)/2+1)}{\Gamma^2(|s_{\pm m}|+1)} M^2\left((|s_{\pm m}|-l)/2, |s_{\pm m}|+1; \frac{\rho_{\pm m}^2}{w_f^2}\right), \quad (14)$$

derivating them upon the radial variables $\rho_{\pm m}$. The transverse intensity profiles of the phase singularity beams are rings with dark cores, whose radii can be found from the equation

$$\left(s_{\pm m} - 2x_{\pm m}\frac{(h-g)}{h}\right) M(g, h; x_{\pm m}) + 2x_{\pm m}^2 \frac{g(h-g)}{h^2(h+1)} M(g+1, h+2; x_{\pm m}) = 0, \quad (15)$$

where we have denoted: $x_{\pm m} = \rho_{\pm m}^2 / w_f^2$; $g = (s_{\pm m} - l)/2$; $h = s_{\pm m} + 1$ (the modulus of $s_{\pm m}$ is considered). The derivation of Equation (15) is explained in more detail in the appendices 1 and 2. Its exact solution is possible only for given values of *m*, *p* and *l*. Considering $x_{\pm m}$ with small value, thus, neglecting the term which contains $x_{\pm m}^2$ in the previous Equation (15), the analytical solution is evaluated as: $x_{\pm m} = \frac{s_{\pm m} h}{2(h-g)}$, or

$$(\rho_{\pm m})_{\max} = w_f \sqrt{\frac{(|l \pm mp|+1)|l \pm mp|}{|l \pm mp|+l+2}}. \quad (16)$$

The comparison between the radii values computed from Equation (16) and those ones read from the graphics of the radial intensity distributions, calculated on base on Equation (14) shows that, there is an excellent agreement for not very high values of the topological charge *mp*. More precise results for the vortex radii can be found by solving numerically the transcendental equation, obtained as a condition the first derivative of the intensity distribution upon the radial coordinate to be equal to zero (Equation (15)).

In the graphs in Figures 3, 4 and 5 the normalized radial intensity distributions in the focal plane of a convergent lens with focal length *f*=30 cm, for given *l*, *m* and *p* values are plotted based on the intensity distributions (14). The used parameters are: $w_0$=1 mm and $\lambda$=800 nm.

In Figure 3 the radial intensity distributions in the first diffraction order, for a given value of *p*, and different values of the topological charge *s* (which is varying due to the change of the incident beam azimuthal mode number *l*) are plotted. It can be seen that, for an incident beam with bigger topological charge *l*, the radius of the vortex is increasing, but the maximum intensity value in the bright ring, surrounding the core, is decreasing.

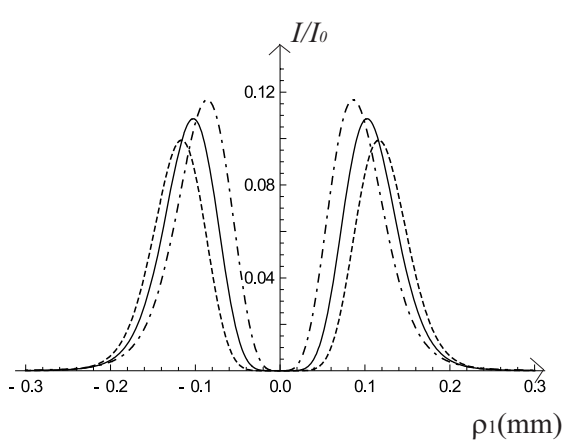
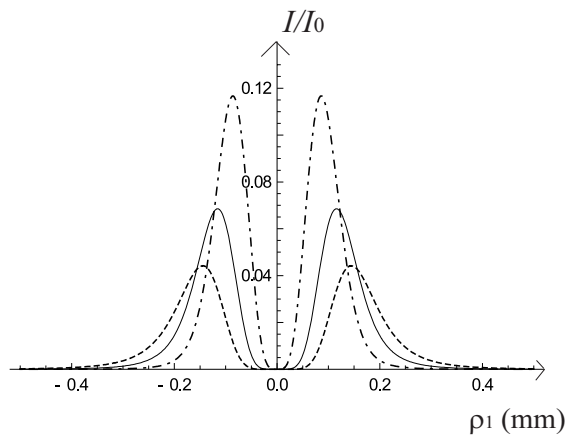

Figure 3. Radial intensity profiles of the diffracted beams in the first diffraction order ($m=1$) for $p=1$, and:
$l=1$ (dot-dashed curve), $l=2$ (solid curve), $l=3$ (dashed curve).

Figure 4. Radial intensity profiles of the diffracted beams in the first diffraction order ($m=1$) for $l=1$, and:
$p=1$ (dot-dashed curve), $p=2$ (solid curve), $p=3$ (dashed curve).

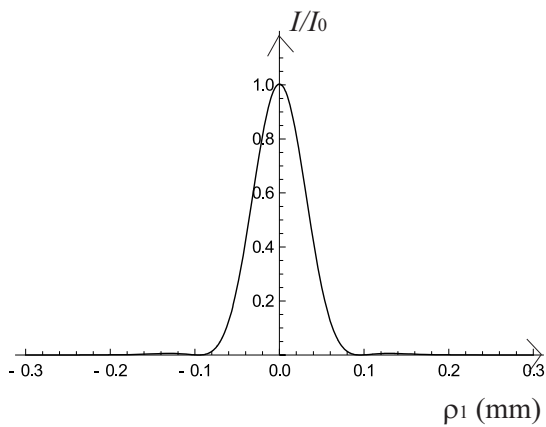

Figure 5. Radial intensity profile of the diffracted beam in the first diffraction order ($m=1$) for $p=-1$, $l=1$.

Also, if the topological charge $s$ has been increased, but, now by increasing the grating's topological charge $p$ (Figure 4), the radius of the dark spot and the radial distance of the maximum intensity value are increasing. The intensity is decreasing more abruptly on the internal side of the vortex, compared to the external side.

In Figure 5 the radial intensity distribution in the first diffraction order for $l=1$, $p=-1$ is shown, in order to see that, when the output beam topological charge ($s=l-mp$) is equal to zero, the beam has a nonzeroth intensity value ($I_0$) along its propagation axis.

### *Specialization of the results when p=0*

When the $\text{LG}_{n=0}^{(l)}$ beam is diffracted by a rectilinear grating ($p=0$), then, in the higher diffraction orders the wave amplitudes are

$$U_{\pm m}(\rho_{\pm m},\theta_{\pm m},z)=iA_{l,0}t_{\pm m}2^{l/2}\frac{w_0}{w_f}\left[\frac{\rho_{\pm m}}{w_f}\right]^{|l|}\exp[il(\pi/2+\theta_{\pm m})]\exp\left(-\frac{\rho_{\pm m}^2}{w_f^2}\right), \qquad (17)$$

while, the zeroth-diffraction-order beam is equal to the reduced in amplitude (by $t_0$) incident beam. The beams diffracted in all diffraction orders have equal topological charges, $l$, which are same as that of the incident beam. They, also, have equal vortex radii: $\rho_0 = \rho_{\pm m} = w_f\sqrt{l/2}$. This specialization agrees with the results reported in [27].

*Specialization of the results when l=0*

For the case of diffraction of a Gaussian beam by fork-shaped grating, from Equation (7), by substitution $l=0$ we get expressions where the Kummer function can be expressed through the difference of two modified Bessel functions (see the identity (A6) in [17]). Thus, we arrive at the final result equal to Equation (42) in [17]. Whereas, from Equation (6) for $l=0$ we get the zeroth-diffraction-order beam as an ordinary Gaussian beam as that one given by Equation (41) in [17].

For the vortex radii, from Equation (16), substituting $l=0$, the following expression is obtained

$$(\rho_{\pm m})_{\max} = w_f\sqrt{\frac{(mp+1)mp}{mp+2}}. \qquad (18)$$

It differs from Equation (46) in [17] by a multiplicator $\sqrt{2}$, because there, the expression for the vortex radii was derived by introducing an approximate, paraxial formula for the modified Bessel function of small argument. Whereas, in this article the vortex radii expression is derived from the intensity distribution expressed through the Kummer function, without applying the Bessel beam approximation. The expression (18) is considered as a more exact. Anyway, in both of them the approximation of neglecting the quadratic terms $\left(\rho_{\pm m}^2/w_f^2\right)^2$ was done, so, these formulas are more valid for not very high values of the topological charge $mp$. More precise results for the vortex radii can be found by solving numerically the equations obtained as a condition the first derivative of the intensity distributions upon the radial coordinate to be equal to zero.

It is also interesting to point out that, in the far-field approximation the derived expression for the vortex radii is same as that given by Equation (16), but, instead of the multiplicator $w_f$ there is $w(z)$ (the beam transverse amplitude profile radius for the fundamental mode, a distance $z$ from its beam waist).

### 4. Conclusions

Summarizing, we have obtained an analytical expression for the paraxial solution of Fraunhofer diffraction of a Laguerre-Gaussian beam with zeroth radial mode number and arbitrary azimuthal mode number $l$, by a diffraction grating with integer forked dislocations $p$. It shows similarity with the far-field approximation solution for Fresnel diffraction of $LG_{n=0}^{(l)}$ beam by the forked grating, which has been also discused. After, the change of the diffracted beams topological charge for different correlation between *l* and *p* has been analyzed. In addition we derived analytical formulas for the vortex radii in the Fourier plane. The results obtained are being specialized for the cases when *p*=0, and when the azimuthal mode number is equal to zero (*l*=0). Each diffracted beam can be transformed by using the same hologram into a helical mode with specific, prescribed topological charge by changing the incident beam topological charge.

Two near-by optical vortices having opposite topological charges exert torques with opposite directions that together can be used to create a microfluidic pump. The resulting fluid flows can be reconfigured dynamically by changing the topological charges, intensities and positions of the optical vortices in an array, which could be potentially useful for microfluids and lab-on-a-chip applications [28].

**Appendix 1**

The intensity in the higher-diffraction-order beams depends upon the radial coordinate, according to Equation (14) as

$$I \propto \exp(-2x_{\pm m}) x_{\pm m}^{s_{\pm m}} M^2(g, h; x_{\pm m}), \quad (A.1.1)$$

where the following notations are used

$$x_{\pm m} = \frac{\rho_{\pm m}^2}{w_f^2}; \quad g = \frac{s_{\pm m} - l}{2}; \quad h = s_{\pm m} + 1. \quad (A.1.2)$$

We are interested in the first derivative of the intensity (A.1.1) upon the radial coordinate

$$\frac{dI}{d\rho_{\pm m}} = \frac{dI}{dx_{\pm m}} \frac{dx_{\pm m}}{d\rho_{\pm m}} = \frac{2\rho_{\pm m}}{w_f^2} \frac{dI}{dx_{\pm m}}, \quad (A.1.3)$$

taking into consideration that

$$\frac{dI}{dx_{\pm m}} = \exp(-2x_{\pm m}) x_{\pm m}^{(s_{\pm m}-1)} M(g, h; x_{\pm m}) \left[ s_{\pm m} M(g, h; x_{\pm m}) + 2x_{\pm m} \frac{dM(g, h; x_{\pm m})}{dx_{\pm m}} - 2x_{\pm m} M(g, h; x_{\pm m}) \right]. \quad (A.1.4)$$

After replacing (A.1.4) into (A.1.3) we search for the root of Equation (A.1.3). Except in the vortex centre $\rho_{\pm m} = 0$, this derivative is also annulated when being satisfied the following identity

$$s_{\pm m} M(g, h; x_{\pm m}) + 2x_{\pm m} \frac{dM(g, h; x_{\pm m})}{dx_{\pm m}} - 2x_{\pm m} M(g, h; x_{\pm m}) = 0. \quad (A.1.5)$$

Further, in (A.1.5) the following relation has been applied [26]

$$\frac{dM(g, h; x_{\pm m})}{dx_{\pm m}} = M(g, h; x_{\pm m}) - \frac{(h-g)}{h} M(g, h+1; x_{\pm m}), \quad (A.1.6)$$

after which we obtain

$$sM(g, h; x_{\pm m}) - 2x_{\pm m} \frac{(h-g)}{h} M(g, h+1; x_{\pm m}) = 0. \quad (A.1.7)$$

Following the definition of the Kummer function we show that it is valid (see appendix 2)

$$M(g, h+1; x_{\pm m}) = M(g, h; x_{\pm m}) - \frac{x_{\pm m}}{h} \frac{g}{(h+1)} M(g+1, h+2; x_{\pm m}). \quad (A.1.8)$$

This, after being involved in (A.1.7), leads to the final equation

$$\left(s_{\pm m} - 2x_{\pm m} \frac{(h-g)}{h}\right) M(g, h; x_{\pm m}) + 2x_{\pm m}^2 \frac{g(h-g)}{h^2(h+1)} M(g+1, h+2; x_{\pm m}) = 0. \quad (A.1.9)$$

**Appendix 2 (derivation of Equation (A.1.8))**

We start from the definitions of the confluent hypergeometric or Kummer functions and the Pochhammer symbols

$$M(g, h; x) = \sum_{\mu=0}^{\infty} \frac{(g)_\mu}{(h)_\mu} \frac{x^\mu}{\mu!}; \quad (g)_\mu = \frac{\Gamma(g+\mu)}{\Gamma(g)}.$$

Further, the Kummer function $M(g, h+1; x)$ is transformed into the following way

$$M(g, h+1; x) = \sum_{\mu=0}^{\infty} \frac{(g)_\mu}{(h+1)_\mu} \frac{x^\mu}{\mu!} = \sum_{\mu=0}^{\infty} \frac{(h+\mu-\mu)}{(h+\mu)} \frac{(g)_\mu}{(h)_\mu} \frac{x^\mu}{\mu!} = \sum_{\mu=0}^{\infty} \frac{(g)_\mu}{(h)_\mu} \frac{x^\mu}{\mu!} - \sum_{\mu=0}^{\infty} \frac{\mu}{(h+\mu)} \frac{(g)_\mu}{(h)_\mu} \frac{x^\mu}{\mu!}$$

$$= M(g, h; x) - \sum_{\mu=0}^{\infty} \frac{\mu}{(h+\mu)} \frac{(g)_\mu}{(h)_\mu} \frac{x^\mu}{\mu!} = M(g, h; x) - \sum_{\mu=1}^{\infty} \frac{1}{(h+\mu)} \frac{(g)_\mu}{(h)_\mu} \frac{x^\mu}{(\mu-1)!},$$

(A.2.1)

where, we have applied the equation

$$(h+1)_\mu = \frac{\Gamma(h+1+\mu)}{\Gamma(h+1)} = \frac{(h+\mu)}{h} \frac{\Gamma(h+\mu)}{\Gamma(h)} = \frac{(h+\mu)}{h}(h)_\mu. \quad (A.2.2)$$

From the upper equation the coefficient $(h)_\mu$ is expressed and replaced in (A.2.1), leading to

$$M(g, h+1; x) = M(g, h; x) - \frac{x}{h} \sum_{\mu=1}^{\infty} \frac{(g)_\mu}{(h+1)_\mu} \frac{x^{\mu-1}}{(\mu-1)!} = M(g, h; x) - \frac{x}{h} \sum_{\mu'=0}^{\infty} \frac{(g)_{\mu'+1}}{(h+1)_{\mu'+1}} \frac{x^{\mu'}}{\mu'!}. \quad (A.2.3)$$

Since it is valid: $(g)_{\mu'+1} = \frac{g}{g} \frac{\Gamma(g+\mu'+1)}{\Gamma(g)} = \frac{g\Gamma(g+\mu'+1)}{\Gamma(g+1)} = g(g+1)_{\mu'}$, the previos equation can be rewritten as

$$M(g, h+1; x) = M(g, h; x) - \frac{x}{h} \frac{g}{(h+1)} \sum_{\mu'=0}^{\infty} \frac{(g+1)_{\mu'}}{(h+2)_{\mu'}} \frac{x^{\mu'}}{\mu'!} = M(g, h; x) - \frac{x}{h} \frac{g}{(h+1)} M(g+1, h+2; x).$$

(A.2.4)